\documentclass[conference]{IEEEtran}
\IEEEoverridecommandlockouts

\usepackage{cite}
\usepackage{amsmath,amssymb,amsfonts}
\usepackage{algorithmic}
\usepackage{graphicx}
\usepackage{textcomp}
\usepackage{xcolor}
\usepackage{booktabs} 
\usepackage{hyperref}
\def\BibTeX{{\rm B\kern-.05em{\sc i\kern-.025em b}\kern-.08em
    T\kern-.1667em\lower.7ex\hbox{E}\kern-.125emX}}
\begin{document}

\title{Feature-Aligned Speech Watermarking for Robustness to Reconstruction Distortions}

\author{
\IEEEauthorblockN{
Haiyun Li\textsuperscript{1,2,}\IEEEauthorrefmark{1},
Shuhai Peng\textsuperscript{1,}\IEEEauthorrefmark{1},
Zhisheng Zhang\textsuperscript{1},
Jingran Xie\textsuperscript{1},
Xiaofeng Xie\textsuperscript{3},
Hanyang Peng\textsuperscript{2,}\IEEEauthorrefmark{2},
Zhiyong Wu\textsuperscript{1,}\IEEEauthorrefmark{2}
}
\IEEEauthorblockA{
\textsuperscript{1}Shenzhen International Graduate School, Tsinghua University, China \\
\textsuperscript{2}Pengcheng Laboratory, China, 
\textsuperscript{3}Independent Researcher, China
\thanks{\IEEEauthorrefmark{1}These authors contributed equally.}
\thanks{\IEEEauthorrefmark{2}Corresponding authors.} 
\thanks{Demo and code: \url{https://hyli-research.github.io/AlignMark/}}
}
\IEEEauthorblockA{
\texttt{\{lihaiyun24, psh24, zhangzs25, xjr21\}@mails.tsinghua.edu.cn}, \\
\texttt{xiexiaofeng1926@gmail.com},
\texttt{penghy@pcl.ac.cn}, 
\texttt{zywu@sz.tsinghua.edu.cn}
}
}

\maketitle

\begin{abstract}
Audio watermarking aims to embed identifiable information into audio while remaining imperceptible. Existing methods adopt high-fidelity, low-energy designs to preserve perceptual quality, but the resulting watermarks lack robustness under suppression by speech reconstruction models. Improving robustness is challenging due to the inherent robustness–fidelity trade-off in existing designs, where increasing watermark energy improves robustness but reduces fidelity.
To address this problem, we propose a feature-aligned watermarking method that aligns the watermark with the original speech feature distribution, allowing higher watermark energy to improve robustness while preserving imperceptibility. We use a pretrained speech codec to generate a pseudo-speech watermark and fuse it into the spectrogram of the input audio, with VAD loss and perceptual losses guiding embedding within voiced regions. Experiments show that our method maintains imperceptibility comparable to existing approaches while substantially improving robustness under both seen and unseen speech reconstruction models.
\end{abstract}


\section{Introduction}
\label{sec:intro}

Audio watermarking embeds detectable information into an audio signal while preserving imperceptibility for human listeners and is widely used for content attribution and copyright protection. To ensure imperceptibility, typical approaches use high-fidelity, low-energy designs that constrain watermark energy to be extremely small, making the watermarked audio highly consistent with the original. This design paradigm is adopted by most current audio watermarking methods.

\begin{figure*}[t]
    \centering
    \includegraphics[width=\linewidth]{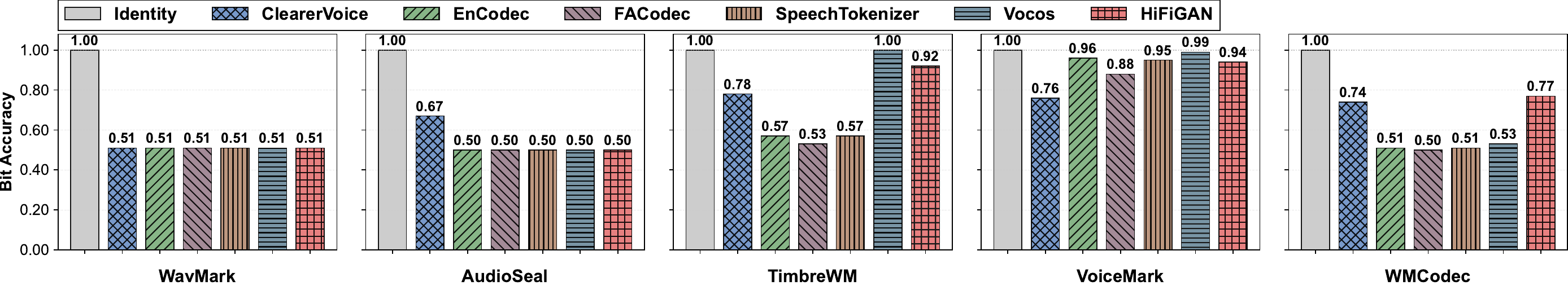}
    \caption{
        Impact of speech reconstruction models on the decoding bit accuracy of existing audio watermarking methods.
        Applying denoising (ClearerVoice \cite{zhao2025clearervoice}), neural codec (EnCodec \cite{defossez2022high}, FACodec \cite{junaturalspeech}), SpeechTokenizer \cite{zhangspeechtokenizer}, and vocoder (HiFiGAN \cite{kong2020hifi}, Vocos \cite{siuzdakvocos}) models leads to varying degrees of degradation across different watermarking methods, even though these reconstruction models are not trained with any watermark-related objective.
    }
    \label{fig:recon_robustness}
\end{figure*}

In modern speech applications such as voice calls, online meetings and social media platforms, speech reconstruction models are widely used. Typical examples include denoising models, neural codecs and vocoders. These models reconstruct clean speech based on the prior speech feature distribution and suppress low-energy noise. This process may also distort or remove the embedded watermark. Several studies \cite{deepwatermarksareshallow, wen2025sok, ozer25_interspeech} have demonstrated that even without any watermark-related knowledge, applying speech reconstruction alone can significantly reduce decoding accuracy. Our experiments similarly confirm this effect, as shown in Fig. \ref{fig:recon_robustness}. This issue significantly limits the applicability of modern audio watermarking methods.

However, addressing this issue is challenging because robustness and fidelity constitute an trade-off in current watermarking methods. Increasing robustness generally requires adding a higher-energy watermark, which inevitably reduces fidelity. This trade-off is inherent to the fundamental design of existing watermarking methods. For example, existing watermarking methods jointly optimize perceptual losses and decoding losses, which inherently conflict with each other. After training convergence, further reducing the decoding loss inevitably increases the perceptual loss. Under this trade-off, typical embedding-based methods, such as WavMark \cite{chen2023wavmark}, AudioSeal \cite{san2024proactive}, and TimbreWM \cite{timbrewatermarking-ndss2024}, will embed very low-energy watermarks into the original audio to maintain fidelity. On the other hand, generative watermarking approaches such as VoiceMark \cite{li25g_interspeech} and WMCodec \cite{zhou2025wmcodec} differ from embedding-based methods as they train generative models to produce watermarked audio directly, offering potential robustness against speech reconstruction. However, beyond addressing the watermarking trade-off, these methods must also jointly optimize the generation task, which substantially increases training complexity, and the resulting audio typically exhibits lower fidelity compared with embedding-based methods. Therefore, within existing architectural designs, improving robustness while maintaining high fidelity under strict watermark-energy constraints presents a significant challenge.

To address this challenge, we take a different perspective. Instead of constraining watermark energy, we align the feature distribution of the watermark more closely with the original speech to maintain imperceptibility for human listeners. Prior work \cite{kong2020hifi, junaturalspeech, zhangspeechtokenizer} has shown that preserving consistency with the original feature distribution is crucial for maintaining perceptual naturalness and intelligibility. Motivated by this, we design a watermark whose feature distribution is aligned with the original speech, enabling it to fuse naturally with the audio and remain imperceptible to human listeners. This design allows the watermark to carry slightly higher energy while still preserving imperceptibility, providing a more flexible trade-off between robustness and fidelity and making it possible to further improve resistance to speech reconstruction.

In this paper, we propose a feature-aligned watermarking method that explicitly aligns the feature distribution of the watermark with the original speech. We adopt a pretrained speech codec model commonly used in speech reconstruction tasks and embed the watermark in its latent space to generate a pseudo-speech watermark that better aligns with the speech features. The pseudo-speech watermark is then fused with the original audio in the spectrogram using learnable fusion weights. We combine a voice activity detection (VAD) loss with perceptual losses such as auditory masking, adversarial, and speaker similarity losses, jointly training the embedder and decoder to concentrate the watermark within voiced regions of the spectrogram, thereby avoiding artifacts and improving imperceptibility. Although this design slightly degrades fidelity due to mixing reconstructed pseudo-speech with the original audio, subjective ABX listening tests \cite{munson1950standardizing} and VISQOL MOS \cite{hines2012visqol} results show that our method achieves imperceptibility comparable to existing embedding-based approaches and substantially better than generative ones. Furthermore, experiments across various speech reconstruction models demonstrate that our approach consistently achieves substantially higher robustness, even against reconstruction models unseen during training. The contributions of this work 
are as follows:
\begin{itemize}
    \item We propose a feature-aligned speech watermarking method that mitigates suppression by speech reconstruction while maintaining imperceptibility to human listeners, thereby improving the applicability of audio watermarking in modern speech applications.
    \item We develop a watermarking framework that embeds the watermark in pretrained codec latents to generate a pseudo-speech watermark, which is fused with the original audio in the spectrogram. A VAD loss and multiple perceptual losses are introduced to align watermark with voiced segments and to maintain imperceptibility.
    \item Experiments show that our method maintains imperceptibility comparable to existing watermarking approaches while achieving superior robustness under both seen and unseen speech reconstruction models, making watermarking more reliable in modern speech applications.
\end{itemize}

\section{Methodology}
\begin{figure*}[t]
\centering
\includegraphics[width=0.8\linewidth]{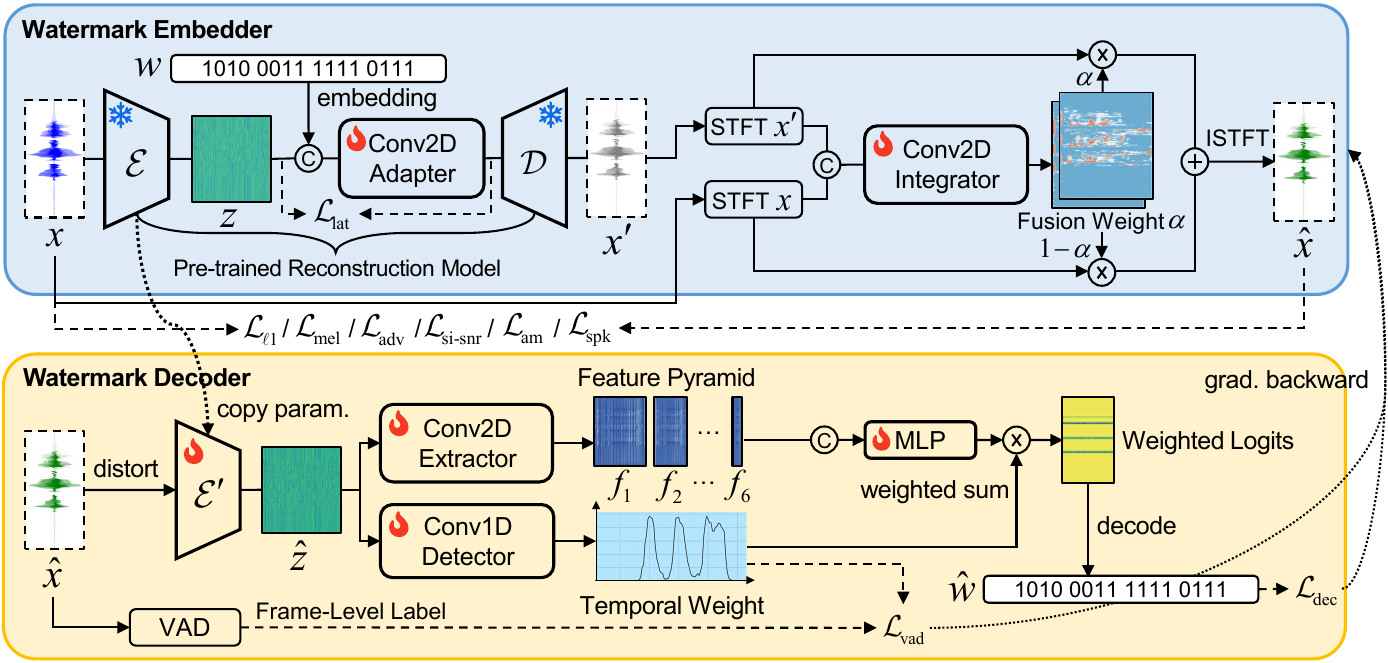}
\caption{The overall architecture of the proposed watermarking framework.}
\label{fig:architecture}
\end{figure*}

The architecture of our method, shown in Figure~\ref{fig:architecture}, consists of a watermark embedder and a watermark decoder. The embedder inserts the watermark into a pretrained codec latent space to generate a pseudo-speech watermark whose feature distribution is aligned with the original speech. This pseudo-speech watermark is then fused with the original audio in the spectrogram domain using learnable fusion weights. VAD loss, decoding loss, and perceptual losses jointly train the embedder and decoder to embed and extract the watermark within voiced regions while improving both robustness and imperceptibility.

\subsection{Watermark Embedder}

The watermark embedder includes a frozen pretrained speech codec encoder–decoder \cite{zhangspeechtokenizer} (without quantization), an adapter $\mathcal{A}$ that injects watermark information into the latent space, and an integrator $\mathcal{I}$ that fuses the pseudo-speech watermark with the original audio in the spectrogram.

\textbf{Pseudo-Speech Generation.}  
Given an input audio waveform $x \in \mathbb{R}^T$, the codec encoder $\mathcal{E}$ extracts latent features $z \in \mathbb{R}^{d \times t}$. To embed an $n$-bit watermark $w \in \{0,1\}^n$, we first map each bit to an embedding vector and sum them to obtain $w_e \in \mathbb{R}^d$. This vector is broadcast across all frames and concatenated with $z$ to form $z_w \in \mathbb{R}^{2d \times t}$. The adapter $\mathcal{A}$, implemented as a 6-layer 2D CNN, transforms $z_w$ into modified latents $z' \in \mathbb{R}^{d \times t}$, which are then input to the codec decoder $\mathcal{D}$ to generate a pseudo-speech watermark audio $x' \in \mathbb{R}^T$. This audio feature closely follows the original speech while carrying the watermark, allowing higher watermark energy to be embedded with minimal perceptible differences.

\textbf{Spectrogram Fusion.}  
To further enhance imperceptibility, we integrate $x'$ into the original audio by converting both $x$ and $x'$ into complex spectrograms $\mathbf{s}_x, \mathbf{s}_{x'} \in \mathbb{R}^{2f \times l}$ via STFT. Their concatenation $\mathbf{s}_c \in \mathbb{R}^{4f \times l}$ is processed by a 4-layer 2D CNN integrator to predict frame–frequency fusion weights $\alpha \in \mathbb{R}^{2f \times l}$. The fused spectrogram is computed by $\mathbf{s}_{w} = \mathbf{s}_{x'} \cdot \alpha + \mathbf{s}_{x} \cdot (1 - \alpha)$. The inverse STFT produces the final watermarked audio $\hat{x}$.

The pseudo-speech design first aligns the feature distribution of the watermark with the original speech. We further improve imperceptibility through the VAD loss and perceptual losses such as auditory masking and speaker similarity. The VAD loss provides temporal guidance, and the perceptual losses provide spectral guidance, jointly encouraging the integrator to fuse the watermark into voiced regions of the spectrogram. This allows the watermark to remain imperceptible while carrying higher energy, improving robustness against speech reconstruction.

\subsection{Watermark Decoder}

The watermark decoder contains a temporal-weight detector and a feature-pyramid extractor. The detector predicts temporal weights that indicate the frame-wise likelihood of voiced segments, while the extractor adopts a multi-scale design to recover the embedded watermark from the speech features.

During training, the watermarked audio $\hat{x}$ is passed through differentiable distortions to produce diverse distorted audio $\tilde{x}$ for robust training. Following prior work \cite{san2024proactive,li25g_interspeech,timbrewatermarking-ndss2024}, we apply augmentations such as replacement, masking, frame shuffle, compression, filtering, and reconstruction through codec \cite{defossez2022high} and vocoder \cite{siuzdakvocos} models. The distorted audio $\tilde{x}$ is then fed into a feature encoder $\mathcal{E}'$ initialized from the frozen codec encoder, producing speech latents $\hat{z} \in \mathbb{R}^{d \times t}$ for subsequent temporal weight prediction and watermark decoding.

\textbf{Temporal Weight Prediction.}  
A 4-layer 1D convolutional detector generates temporal weights $p \in \mathbb{R}^t$ via:
\begin{equation}
p = \text{Sigmoid}(\text{Conv1D}(\hat{z})).
\end{equation}
These weights indicate the frame-wise likelihood of voiced segments, and the VAD loss jointly guides the embedder and decoder to focus the watermark within voiced frames.

\textbf{Feature Pyramid Extraction.}  
Since the watermark is fused with speech features in voiced regions, its extraction becomes more challenging. To capture the fine-grained details needed for decoding, we adopt a feature-pyramid design. A 6-layer 2D CNN extractor processes the latents into multi-scale features. At scale $i$, the feature map is computed as
\begin{equation}
\hat{z}_i = \text{Conv2D}_i(\hat{z}_{i-1}), \qquad \hat{z}_0 = \hat{z},
\end{equation}
where $\hat{z}_{i} \in \mathbb{R}^{c_i \times (d/2^{i}) \times t}$ and $c_i = 2^{i-1} \cdot 16$.

To produce a unified time-aligned representation, each $\hat{z}_i$ is projected along its second dimension (the reduced latent dimension $d/2^i$) using a fully connected layer:
\begin{equation}
f_i = \text{FC}_i(\hat{z}_i) \in \mathbb{R}^{c_i \times t}.
\end{equation}

The final feature pyramid is obtained by concatenating all $f_i$ along the feature dimension, 
$f = \text{Concat}(f_1, f_2, \dots, f_6) \in \mathbb{R}^{c \times t}$, 
where $c = \sum_{i=1}^{6} c_i$.

\textbf{Watermark Decoding.}  
The feature pyramid $f$ is then fed into a 2-layer MLP, which maps it to frame-wise logits $w_f \in \mathbb{R}^{(n/4) \times 16 \times t}$, where the $(n/4) \times 16$ structure corresponds to the hexadecimal representation of the $n$-bit watermark \cite{li25g_interspeech,zhou2025wmcodec}. Temporal weighting aggregates these logits as $\hat{w} = \sum_{t}(w_f \cdot p)$, and an argmax over the $(n/4) \times 16$ logits followed by base conversion yields the final $n$-bit decoded watermark.

The training of the temporal weight allows gradients to propagate to $\mathcal{E}'$ and the entire embedder, ensuring that both the embedder and decoder focus on the same voiced segments. The feature pyramid provides sufficient feature extraction capacity to decode the watermark from speech features, enabling improved robustness while maintaining imperceptibility.

\begin{table*}[t]
\centering
\caption{
Robustness comparison on speech reconstruction (ACC $\uparrow$, FAR $\downarrow$; \textbf{bold} indicates the best). 
Methods marked with $^\dagger$ are our unseen distortions. 
Note that SpeechTokenizer is only used as a pretrained model and is not applied as a distortion during training.
}
\label{tab:robustness_neural}
\begin{tabular}{lcccccccccccc}
    \toprule
    & \multicolumn{2}{c}{WavMark} 
    & \multicolumn{2}{c}{AudioSeal} 
    & \multicolumn{2}{c}{TimbreWM} 
    & \multicolumn{2}{c}{VoiceMark} 
    & \multicolumn{2}{c}{WMCodec} 
    & \multicolumn{2}{c}{Ours} \\
    \cmidrule(lr){2-3} \cmidrule(lr){4-5} \cmidrule(lr){6-7} 
    \cmidrule(lr){8-9} \cmidrule(lr){10-11} \cmidrule(lr){12-13}
    Method & ACC & FAR & ACC & FAR & ACC & FAR & ACC & FAR & ACC & FAR & ACC & FAR \\
    \midrule
    ClearerVoice$^\dagger$      & 0.51 & 1.00 & 0.67 & 0.79 & 0.78 & 0.65 & 0.76 & 0.60 & 0.74 & 0.83 & 0.92 & 0.27 \\
    EnCodec                     & 0.51 & 1.00 & 0.50 & 1.00 & 0.57 & 0.99 & 0.96 & 0.16 & 0.51 & 1.00 & 0.99 & 0.02 \\
    FACodec$^\dagger$           & 0.51 & 1.00 & 0.50 & 0.99 & 0.53 & 1.00 & 0.88 & 0.45 & 0.50 & 1.00 & 0.93 & 0.24 \\
    SpeechTokenizer$^\dagger$   & 0.51 & 1.00 & 0.50 & 0.99 & 0.57 & 0.99 & 0.95 & 0.21 & 0.51 & 1.00 & 0.96 & 0.11 \\
    Vocos                       & 0.51 & 1.00 & 0.50 & 1.00 & 1.00 & 0.00 & 0.99 & 0.04 & 0.53 & 1.00 & 1.00 & 0.00 \\
    HiFiGAN$^\dagger$           & 0.51 & 1.00 & 0.50 & 1.00 & 0.92 & 0.48 & 0.94 & 0.22 & 0.77 & 0.79 & 0.99 & 0.01 \\
    \midrule
    Average                     & 0.51 & 1.00 & 0.53 & 0.96 & 0.73 & 0.69 & 0.91 & 0.28 & 0.59 & 0.94 & \textbf{0.97} & \textbf{0.11} \\
    \bottomrule
\end{tabular}

\end{table*}

\begin{table*}[t]
\centering
\caption{Robustness against traditional audio distortions (ACC $\uparrow$, FAR $\downarrow$; \textbf{bold} indicates the best). }
\label{tab:robustness_traditional}
\begin{tabular}{l cccccccccccc}
    \toprule
    Method 
    & \multicolumn{2}{c}{WavMark} 
    & \multicolumn{2}{c}{AudioSeal} 
    & \multicolumn{2}{c}{TimbreWM} 
    & \multicolumn{2}{c}{VoiceMark} 
    & \multicolumn{2}{c}{WMCodec} 
    & \multicolumn{2}{c}{Ours} \\
    \cmidrule(lr){2-3} \cmidrule(lr){4-5} \cmidrule(lr){6-7} \cmidrule(lr){8-9} \cmidrule(lr){10-11} \cmidrule(lr){12-13}
    & ACC & FAR & ACC & FAR & ACC & FAR & ACC & FAR & ACC & FAR & ACC & FAR \\
    \midrule
    Resampling     & 1.00 & 0.00 & 1.00 & 0.00 & 1.00 & 0.00 & 0.99 & 0.06 & 1.00 & 0.01 & 1.00 & 0.00 \\
    Boost Volume   & 1.00 & 0.00 & 1.00 & 0.00 & 1.00 & 0.00 & 0.98 & 0.09 & 1.00 & 0.01 & 1.00 & 0.00 \\
    Duck Volume    & 1.00 & 0.00 & 1.00 & 0.00 & 1.00 & 0.00 & 0.98 & 0.07 & 1.00 & 0.01 & 1.00 & 0.01 \\
    Highpass       & 1.00 & 0.00 & 1.00 & 0.00 & 1.00 & 0.00 & 0.99 & 0.05 & 1.00 & 0.01 & 1.00 & 0.00 \\
    Lowpass        & 1.00 & 0.00 & 1.00 & 0.00 & 1.00 & 0.02 & 0.76 & 0.82 & 0.85 & 0.67 & 1.00 & 0.00 \\
    Bandpass       & 1.00 & 0.01 & 1.00 & 0.00 & 0.99 & 0.06 & 0.76 & 0.72 & 0.82 & 0.79 & 1.00 & 0.00 \\
    AAC            & 1.00 & 0.00 & 0.73 & 0.96 & 1.00 & 0.00 & 0.98 & 0.07 & 1.00 & 0.01 & 0.99 & 0.02 \\
    MP3            & 0.98 & 0.04 & 1.00 & 0.00 & 1.00 & 0.00 & 0.85 & 0.56 & 0.92 & 0.40 & 0.99 & 0.01 \\
    Echo           & 0.99 & 0.02 & 1.00 & 0.00 & 1.00 & 0.00 & 0.98 & 0.11 & 1.00 & 0.01 & 1.00 & 0.01 \\
    Crop           & 0.98 & 0.03 & 0.62 & 0.91 & 1.00 & 0.00 & 0.98 & 0.08 & 1.00 & 0.01 & 0.99 & 0.02 \\
    Pink Noise     & 0.98 & 0.05 & 1.00 & 0.01 & 1.00 & 0.01 & 0.99 & 0.05 & 1.00 & 0.01 & 1.00 & 0.00 \\
    Gaussian Noise & 0.51 & 1.00 & 0.77 & 0.73 & 0.87 & 0.60 & 0.54 & 0.99 & 0.75 & 0.76 & 0.99 & 0.02 \\
    Smooth         & 0.98 & 0.03 & 1.00 & 0.00 & 1.00 & 0.00 & 0.76 & 0.72 & 0.97 & 0.14 & 0.99 & 0.03 \\
    Pitch Shift    & 0.52 & 0.97 & 0.53 & 0.97 & 0.54 & 0.72 & 0.85 & 0.50 & 0.66 & 0.72 & 1.00 & 0.00 \\
    Speed Change   & 0.51 & 0.99 & 0.50 & 0.99 & 0.50 & 0.98 & 0.53 & 0.95 & 0.51 & 0.97 & 0.94 & 0.21 \\
    \midrule
    Average        & 0.90 & 0.21 & 0.88 & 0.30 & 0.93 & 0.16 & 0.86 & 0.39 & 0.90 & 0.30 & \textbf{0.99} & \textbf{0.02} \\
    \bottomrule
\end{tabular}
\end{table*}

\subsection{Training Loss}

To maintain imperceptibility while improving robustness, we combine decoding loss and perceptual losses for training.

\textbf{VAD loss.}
Temporal weights are supervised using binary VAD labels to encourage watermark embedding and extraction within voiced frames: $\mathcal{L}_{\text{vad}} = -\frac{1}{t} \sum_{i=1}^{t} \big[ v_i \log p_i + (1 - v_i)\log(1 - p_i) \big]$, where $v_i \in \{0,1\}$ is the VAD label using the implementation in \cite{li25g_interspeech} (0 for silent/masked/replaced frames and 1 otherwise), and $p_i$ is the predicted temporal weight.

\textbf{Auditory masking loss.}
To maintain imperceptibility, we incorporate an auditory masking loss $\mathcal{L}_{\text{am}}$ following \cite{san2024proactive}, which evaluates loudness differences between the watermark component and the original signal across time–frequency windows based on psychoacoustic masking principles.

\textbf{Speaker similarity loss.}
To preserve speaker identity after watermark embedding, we use: $\mathcal{L}_{\text{spk}} = 1 - \cos(\text{Emb}(x), \text{Emb}(\hat{x}))$, where $x$ and $\hat{x}$ are the original and watermarked audio, and $\text{Emb}(\cdot)$ extracts speaker embeddings by Resemblyzer\footnote{https://github.com/resemble-ai/Resemblyzer}.

\textbf{Latent similarity loss.}
To constrain deviations in the latent space and keep the pseudo-speech watermark aligned with the original speech, we apply: $\mathcal{L}_{\text{lat}} = 1 - \cos(z, z')$, where $z$ and $z'$ denote the codec latents before and after watermark embedding.

\textbf{Fidelity losses.}
We adopt several fidelity losses \cite{timbrewatermarking-ndss2024,san2024proactive,li25g_interspeech} to constrain the difference between the watermarked audio $\hat{x}$ and the original audio $x$, including L1 loss $\mathcal{L}_{\ell1}$, Mel-spectrogram loss $\mathcal{L}_{\text{mel}}$, adversarial loss $\mathcal{L}_{\text{adv}}$, and SI-SNR loss $\mathcal{L}_{\text{si-snr}}$.

\textbf{Decoding loss.}
Hexadecimal classification is optimized using cross-entropy: $\mathcal{L}_{\text{dec}} = -\frac{1}{n/4} \sum_{j=1}^{n/4} \sum_{k=1}^{16} y_{jk} \log \hat{w}_{jk}$, where $y_{jk}$ and $\hat{w}_{jk}$ are the one-hot label and predicted probability for the $j$-th hexadecimal digit.

\textbf{Total loss.}
The overall objective is a weighted sum:
\begin{equation}
\mathcal{L}_{\text{total}} = \sum_{\ell \in \mathcal{L}} \lambda_{\ell}\mathcal{L}_{\ell}.
\end{equation}
The weights are set as
$\lambda_{\text{vad}} = 1.0$, 
$\lambda_{\text{am}} = 0.1$, 
$\lambda_{\text{spk}} = 0.1$, 
$\lambda_{\text{lat}} = 0.1$, 
$\lambda_{\ell1} = 0.01$, 
$\lambda_{\text{mel}} = 0.1$, 
$\lambda_{\text{adv}} = 0.5$, 
$\lambda_{\text{si-snr}} = 0.01$.
$\lambda_{\text{dec}} = 4.0$, 
With our pseudo-speech watermark fusion design, small weights can be applied to signal-level losses such as $\mathcal{L}_{\ell1}$, $\mathcal{L}_{\text{mel}}$, and $\mathcal{L}_{\text{si-snr}}$, relaxing fidelity constraints while maintaining imperceptibility and improving robustness.

\section{Experiments}

\subsection{Experimental Setups}
\textbf{Training}.
The codec model uses weights from SpeechTokenizer \cite{zhangspeechtokenizer}. The adapter $\mathcal{A}$ employs skip-gated blocks \cite{timbrewatermarking-ndss2024} with layers of 32 channels. The integrator $\mathcal{I}$ consists of STFT (256 FFT points, hop length 64, window length 256) and 2D convolutions (64 channels, LeakyReLU with slope 0.1). The watermark detector and extractor employ different architectures: the detector uses 1D convolutions (256 channels, GELU), while the extractor uses 2D convolutions (kernel $(5, 3)$, stride $(2, 1)$, padding $(0, 1)$, channels doubling from 16, GELU). The watermark bit length $n$ is 16. Adam optimizer is used with a $5e^{-5}$ learning rate, trained for 300 epochs, selecting the checkpoint with the lowest loss.

\textbf{Dataset}.
Our experiments use three datasets: VCTK \cite{Yamagishi2012}, LibriSpeech \cite{panayotov2015librispeech}, and LJSpeech \cite{ljspeech17}. For VCTK, 200 audio samples are randomly selected for testing, with the remaining samples used for training. Similarly, 200 audio samples are randomly selected from LibriSpeech and LJSpeech respectively, forming a 600-sample test set with a total duration of approximately one hour and without length truncation.

\textbf{Metrics}.
We evaluate robustness using bit-wise accuracy (ACC) and false attribution rate (FAR), following prior work \cite{timbrewatermarking-ndss2024,san2024proactive,li25g_interspeech}. ACC measures the ratio of correctly decoded bits. FAR is computed as the proportion of decoded watermarks whose ground-truth watermark is not the closest match in the test set under Hamming distance. For subjective evaluation, we perform ABX tests \cite{munson1950standardizing} with 20 subjects and 10 samples per method. Listeners compare A and B, then identify whether X (randomly drawn from A or B) matches either; accuracy near 50\% implies imperceptibility. Compared to MOS, ABX more directly captures perceptual differences between original and watermarked audio, making it better suited for watermarking evaluation. We further use the VISQOL MOS objective listening quality metric as a supplementary indicator. VISQOL uses a neural network to evaluate the similarity between the original and watermarked audio on a 1–5 scale, where higher scores indicate better quality. Objective metrics include NISQA \cite{mittag2021nisqa} for assessing perceptual naturalness, as well as standard fidelity metrics \cite{san2024proactive} such as perceptual evaluation of speech quality (PESQ), scale-invariant signal-to-noise ratio (SI-SNR), and short-time objective intelligibility (STOI).

\textbf{Watermarking Methods}
We compare with five state-of-the-art methods with publicly available implementations: WavMark \cite{chen2023wavmark}, AudioSeal \cite{san2024proactive}, TimbreWM \cite{timbrewatermarking-ndss2024}, VoiceMark \cite{li25g_interspeech}, and WMCodec \cite{zhou2025wmcodec}. We use a consistent 16-bit setting for all methods except TimbreWM, which does not provide a 16-bit model; therefore, we use its closest 10-bit version. For WavMark, undetected watermarks are treated as zeros.

\textbf{Distortions}.
We evaluate robustness across 21 distortions, including six speech reconstruction models: a denoiser (ClearerVoice \cite{zhao2025clearervoice}), neural codecs (EnCodec \cite{defossez2022high}, FACodec \cite{junaturalspeech}), SpeechTokenizer \cite{zhangspeechtokenizer}, and vocoders (HiFiGAN \cite{kong2020hifi}, Vocos \cite{siuzdakvocos}). Following prior work \cite{deepwatermarksareshallow}, we add 10 dB Gaussian noise before denoising. We also include 15 traditional distortions using implementations from \cite{san2024proactive,wen2025sok,deepwatermarksareshallow}.

\begin{figure}[t]
\centering
\includegraphics[width=0.9\linewidth]{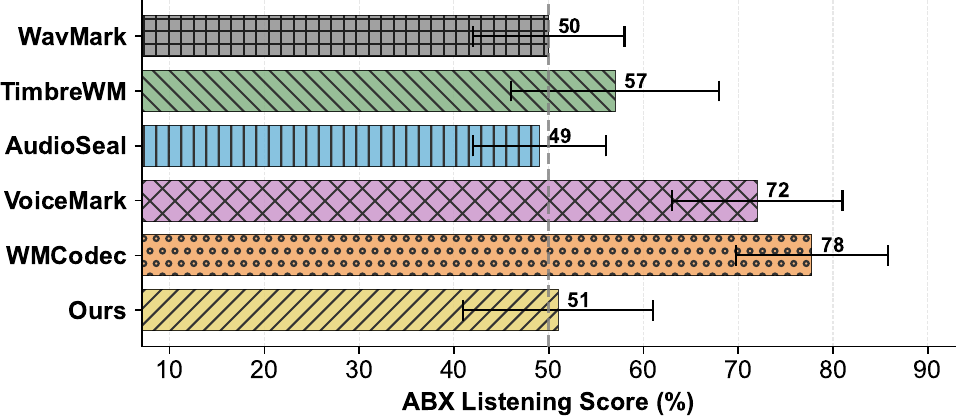}
\caption{
Subjective ABX listening test results measuring the imperceptibility of audio watermarking methods. 
Scores closer to \textbf{50\%} indicate higher imperceptibility, with error bars showing 95\% confidence intervals.
}
\label{fig:abx_boxplot}
\end{figure}

\subsection{Robustness Evaluation}
We evaluate the robustness of our method under two categories of distortions: speech reconstruction models and traditional signal-level distortions. As shown in Table~\ref{tab:robustness_neural}, our method consistently achieves higher ACC and lower FAR than prior watermarking approaches under speech reconstruction models, where many existing methods degrade substantially. For traditional distortions, the results in Table~\ref{tab:robustness_traditional} show that our method performs comparably to prior work in most conditions, with improvements under pitch and speed changes as well as additive noise. Slight drops occur under severe compression or cropping, which may remove portions of the speech that contain the embedded watermark. Overall, the results indicate that our approach improves robustness to reconstruction while maintaining competitive performance under other distortions.

\subsection{Perceptual Audio Quality}

\textbf{Subjective ABX listening test.}
As shown in Figure~\ref{fig:abx_boxplot}, our method achieves an ABX score comparable to embedding-based watermarking methods such as WavMark and AudioSeal, indicating that the perceptual difference between the original and watermarked audio remains close to imperceptible. Generative watermarking approaches such as VoiceMark exhibit larger deviations than others. These results suggest that our method maintains imperceptibility for human listeners.

\begin{figure}[t]
\centering
\includegraphics[width=0.9\linewidth]{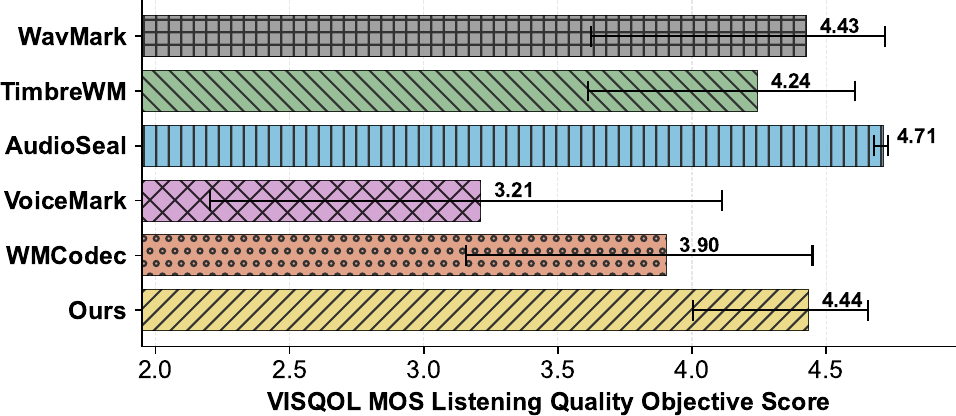}
\caption{
VISQOL MOS listening quality objective scores for different audio watermarking methods. 
Higher scores indicate better perceptual quality. 
}
\label{fig:nisqa_bars}
\end{figure}

\begin{table}[t]
\centering
\caption{Objective audio quality evaluation.}
\label{fig:quality}
\begin{tabular}{lrrrr}
\toprule
Method & NISQA $\uparrow$ & PESQ $\uparrow$ & STOI $\uparrow$ & SI-SNR $\uparrow$ \\
\midrule
WavMark        & 4.21 & 4.10 & 0.99 & 36.87 \\
TimbreWM       & 4.22 & 3.72 & 0.99 & 23.93 \\
AudioSeal      & 4.28 & 4.37 & 0.99 & 27.60 \\
VoiceMark      & 4.36 & 2.19 & 0.90 & 1.90  \\
WMCodec        & 4.22 & 2.31 & 0.91 & -0.62  \\
\midrule
Ours           & 4.31 & 3.03 & 0.95 & 12.16 \\
\bottomrule
\end{tabular}
\end{table}

\textbf{Objective audio quality.}
Figure~\ref{fig:nisqa_bars} reports the VISQOL MOS listening quality scores across different watermarking methods. Our method attains a score of 4.44, ranking second among all baselines and only slightly below AudioSeal. These results indicate that, in terms of perceptual features extracted by the neural network, the watermarked audio remains highly similar to the original audio. Table~\ref{fig:quality} summarizes the naturalness metric (NISQA) and the fidelity metrics (PESQ, STOI, SI-SNR). While our method is not designed to optimize fidelity, it still outperforms generative watermarking methods such as VoiceMark and WMCodec on this metric. These results suggest that our feature-aligned watermark design effectively maintains imperceptibility.


\begin{figure}[htbp]
    \centering
    \includegraphics[width=\linewidth]{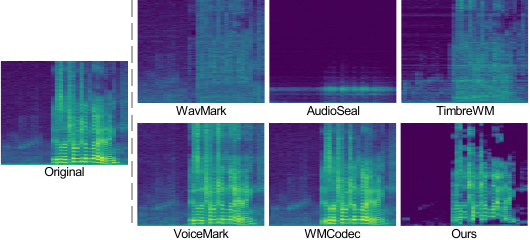}
    \caption{
    Spectrogram comparison for the case study. The first panel shows the original audio; the others show spectrograms of difference signals (watermarked waveform minus original waveform).
    }
    \label{fig:case_study_spec}
\end{figure}

\subsection{Case Study}

Figure~\ref{fig:case_study_spec} compares difference spectrograms obtained by subtracting each watermarked audio from the original. All spectrograms are normalized to facilitate energy comparison. Our method exhibits higher watermark energy than embedding-based approaches while remaining more concentrated and better aligned with speech features than generative methods, with most energy located in voiced regions. This pattern indicates that our watermark can use higher energy to improve robustness while still preserving imperceptibility.

\begin{table}[htbp]
\centering
\caption{Ablation study.}
\label{tab:ablation}
\begin{tabular}{lcccc}
\toprule
Method & ACC $\uparrow$ & FAR $\downarrow$ & PESQ $\uparrow$ & STOI $\uparrow$ \\
\midrule
Full Model        & 0.98 & 0.05 & 3.03 & 0.95 \\
\hspace{1em} w/o Spectrogram Fusion     & 0.80 & 0.49 & 2.75 & 0.94 \\
\hspace{1em} w/o VAD Loss & 0.84 & 0.40 & 1.89 & 0.85 \\
\hspace{1em} w/o Feature Pyramid  & 0.76 & 0.73 & 3.15 & 0.95 \\
\bottomrule
\end{tabular}
\end{table}

\subsection{Ablation Study}

Table~\ref{tab:ablation} shows the ablation results averaged over all distortions. Removing fusion or the VAD loss leads to clear drops, indicating that aligning the watermark with speech features helps resist distortions, including speech reconstruction. Removing the feature pyramid results in the largest drop, suggesting that sufficient feature extraction capacity is necessary for decoding watermarks fused into speech features.

\section{Conclusion}

We propose a feature-aligned watermarking method that aligns the watermark with the original speech feature distribution, allowing higher watermark energy to improve robustness while preserving imperceptibility. The method uses a pretrained speech codec to generate a pseudo-speech watermark and fuses it into the voiced regions of the spectrogram guided by VAD and perceptual losses to improve imperceptibility. Experiments show that our approach maintains imperceptibility comparable to existing methods while achieving substantially higher robustness under various speech reconstruction models.

\section*{Acknowledgment}
This work is supported by National Natural Science Foundation of China (62076144) and Shenzhen Science and Technology Program (JCYJ20220818101014030).

\bibliographystyle{IEEEtran}
\bibliography{IEEEabrv,icme2026references}

\end{document}